\documentclass[aps,prb,twocolumn,superscriptaddress,showpacs]{revtex4}
\usepackage{ulem}
\usepackage[utf8]{inputenc}
\usepackage{graphicx}
\usepackage{dcolumn}
\usepackage{bm}
\usepackage{hyperref}
\usepackage{amsmath}
\usepackage{mathtools}
\usepackage[dvipsnames]{xcolor}

\begin{document}

\preprint{APS/123-QED}

\title{Hyperuniformity in Type-II Superconductors with Point and Planar Defects}

\author{Joaqu\'{i}n Puig, Jazm\'{i}n Arag\'{o}n S\'{a}nchez, Gladys Nieva, Alejandro B. Kolton and Yanina Fasano$^{*}$}

\affiliation{Centro At\'{o}mico Bariloche and Instituto Balseiro,
CNEA, CONICET and Universidad Nacional de Cuyo, 8400 San Carlos de
Bariloche, Argentina}


\date{today}

\begin{abstract}
We use vortex matter in type-II superconductors as a playground to
study how different types of disorder affect the long wavelength
density fluctuations of the system. We find that irrespective of the
vortex-vortex interaction, in the case of samples with  weak and
dense point defects the system presents the hidden order of
hyperuniformity characterized by an algebraic suppression of density
fluctuations when increasing the system size. We also reveal that,
on the contrary, for samples with planar defects hyperuniformity is
suppressed since density fluctuations have a tendency to
unboundedness on increasing the system size. Although some of these
results were known from previous works, this paper makes the
fundamental discovery that the ability of planar disorder to
suppress hyperuniformity grows on increasing the softness of the
structure for more diluted systems.

\end{abstract}

\maketitle

\section{Introduction}

Vortex matter in type-II superconducting samples is a system of
interacting elastic objects that can be used as a laboratory
playground to study the growth of hyperuniform materials nucleated
in host media with disorder.~\cite{Rumi2019,Llorens2020b} The quest
for hyperuniform material systems is currently attracting great
interest in the condensed matter and materials science communities
due to their unique physical properties. Hyperuniform materials are
endowed with a novel phenomenology that goes against the
conventional wisdom on the effect of structural disorder in systems
of interacting objects.~\cite{Man2013,Chen2018,Torquato2018} For
instance, disordered hyperuniform two-dimensional silica structures
present a closing of bandgaps for electrical transport resulting in
an enhanced conductivity.~\cite{Zheng2020} Also, disordered
hyperuniform materials posses complete photonic bandgaps blocking
all directions and polarizations for short
wavelengths,~\cite{Florescu2009,Man2013,FroufePerez2016} in contrast
to previous assumptions that periodic or quasiperiodic order  was a
prerequisite for a material to present this optical property.

Hyperuniformity is a topological property of a state of matter that
is characterized by strongly-reduced long-wavelength density
fluctuations entailing a decaying structure factor $S(\mathbf{q})$
for small wave-vectors $\mathbf{q}$. The density of constituents in
hyperuniform systems is homogeneous at large scales, as in a perfect
lattice, but it can present fluctuations at short length scales as
in a disordered structure.~\cite{Torquato2018,Torquato2003}
Hyperuniformity is a structural property defined in an asymptotic
limit and ascertaining this property in real systems is thus
difficult. For this reason most works show that the systems are
\textit{effectively} hyperuniform.~\cite{Klatt2019}

The structure factor can be directly measured using different X-ray
and neutron diffraction techniques depending on the typical lattice
spacing of the systems. An alternative way of obtaining this
magnitude is to compute it from the real-space positions of the
individual constituents considering that $S(\mathbf{q})=
|\hat{\rho}(q_{\rm x},q_{\rm y})| ^{2}$, with $\hat{\rho}$ the
Fourier transform  of the local density modulation $\rho$. In this
work we use this approach to study the occurrence of hyperuniformity
in vortex matter nucleated in superconducting samples with point and
planar crystal defects. The nature of disorder unavoidably present
in the host medium  affects whether the nucleated structure is
hyperuniform or non-hyperuniform. In the case of samples with weak
and dense point disorder, some of us revealed that the vortex
structure nucleated in the cuprate superconductor
Bi$_{2}$Sr$_{2}$CaCu$_{2}$O$_{8 + \delta}$ is effectively
hyperuniform at the sample surface.~\cite{Rumi2019} Later studies
report that disordered hyperuniform vortex structures are nucleated
in pnictide~\cite{Llorens2020b} and Fe-based~\cite{Aragon2022}
superconductors with point disorder. Nevertheless, we recently found
that the presence of planar correlated defects suppress
hyperuniformity in the vortex structure in an anisotropic
fashion.~\cite{Puig2021} Thus, attention to the nature of disorder
in the host medium has to be payed  in the search for novel
hyperuniform materials composed of interacting objects.

Another relevant parameter that enters into play when trying to
nucleate hyperuniform structures is the magnitude of the interaction
between constituents that when enhanced will tend to decrease the
density fluctuations irrespective of the nature of disorder in the
medium. In our case, the vortex density is controlled by the applied
field since the lattice spacing $a_{0} \propto (B)^{-1/2}$ and, in
addition, the interaction between vortices becomes larger on
increasing field. In this work we study how a softening of the
vortex structure affects the density fluctuations at long
wavelengths or short wavevectors for point and correlated disorder
in the samples.

\section{Experimental}

We image the structure with single vortex resolution at the surface
of the samples by means of the magnetic decoration
experiments~\cite{Fasano2008} performed at 4.2\,K after following a
field-cooling process.\cite{Fasano2003} As demonstrated previously,
during this cooling process the vortex structure gets frozen  at
length-scales of the lattice spacing at a temperature
$T_{\mathrm{freez}} \sim T_{\rm irr}$, the irreversibility
temperature at which pinning sets in on
cooling.~\cite{CejasBolecek2016}  At this crossover temperature the
bulk pinning dominates over the vortex-vortex repulsion and the
thermal fluctuations.~\cite{Pardo1997,Fasano1999} Decoration of
individual vortices is performed at 4.2\,K by evaporating Fe
clusters in a controlled-pressure helium atmosphere. These
magnetized clusters are attracted towards the magnetic halo of the
vortex core due to the magnetic force exerted by the local field
gradient inherent to vortices. Once the experiment is performed the
sample is warmed up to room temperature and the Fe clusters that
remain attached to the surface due to van der Waals forces are
observed by means of scanning electron microscopy. Vortices are
decorated with the Fe clumps and imaged as white dots in the images.

The superconducting samples we study are nearly optimally-doped
single crystals of Bi$_2$Sr$_2$CaCu$_2$O$_{8+\delta}$ ($T_{\rm c}
\sim 90$\,K) grown by means of the flux method.~\cite{Correa2001}.
Some of the crystals present few planar crystallographic defects
consisting in stacking faults in the crystals separating zones with
slightly different orientations of their
$c$-axis.~\cite{Koblischka1995,Herbsommer2001} We also study samples
with point disorder only, namely with no planar defects as revealed
by means of magnetic-decoration imaging of the vortex structure.

\begin{figure}[htb]
\centering
\includegraphics[width=0.8\textwidth]{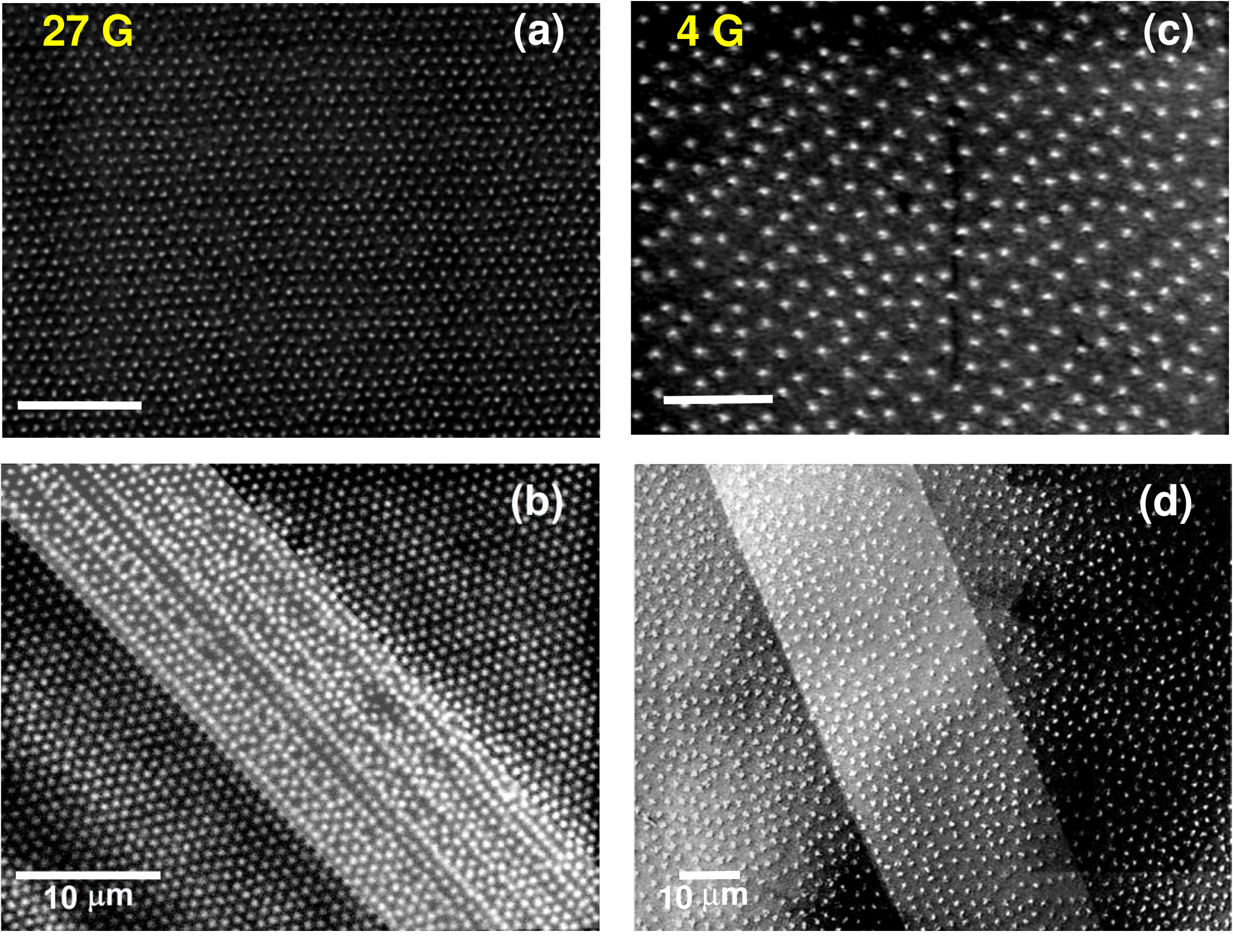}
\caption{Magnetic decoration images of the vortex structure (white
dots) nucleated in Bi$_2$Sr$_2$CaCu$_2$O$_{8+\delta}$ samples with
point (a),(c) and planar (b),(d) disorder. Vortex densities
correspond to (a),(b) 27 and (c),(d) 4\,G. Planar defects are
preferential pinning sites and thus cage vortices in vortex rows,
see highlighted regions in panels (b) and (d). All white scale bars
correspond to 10\,$\mu$m.} \label{fig:Figure1}
\end{figure}

\section{Results}

Figure\,\ref{fig:Figure1} shows snapshots of vortex structures
nucleated in samples with point (top panels) and planar (bottom
panels) defects for two different vortex densities. Data in the left
panels correspond to the vortex structure nucleated at 27\,G and on
the right panels to the softer vortex structure nucleated at 4\,G.
In the case of samples with point defects, the more diluted 4\,G
vortex structure presents a larger amount of density fluctuations at
short lengthscales than the one nucleated at 27\,G. In samples with
planar defects, they are effective to induce vortex rows aligned
along the defects irrespective of the vortex density, see
highlighted regions in the Figs.\,\ref{fig:Figure1} (b) and (d). In
the region of these rows, the vortex structure present strong
density fluctuations for both magnetic fields.

\begin{figure}[htb]
    \centering
    \includegraphics[width=0.6\textwidth]{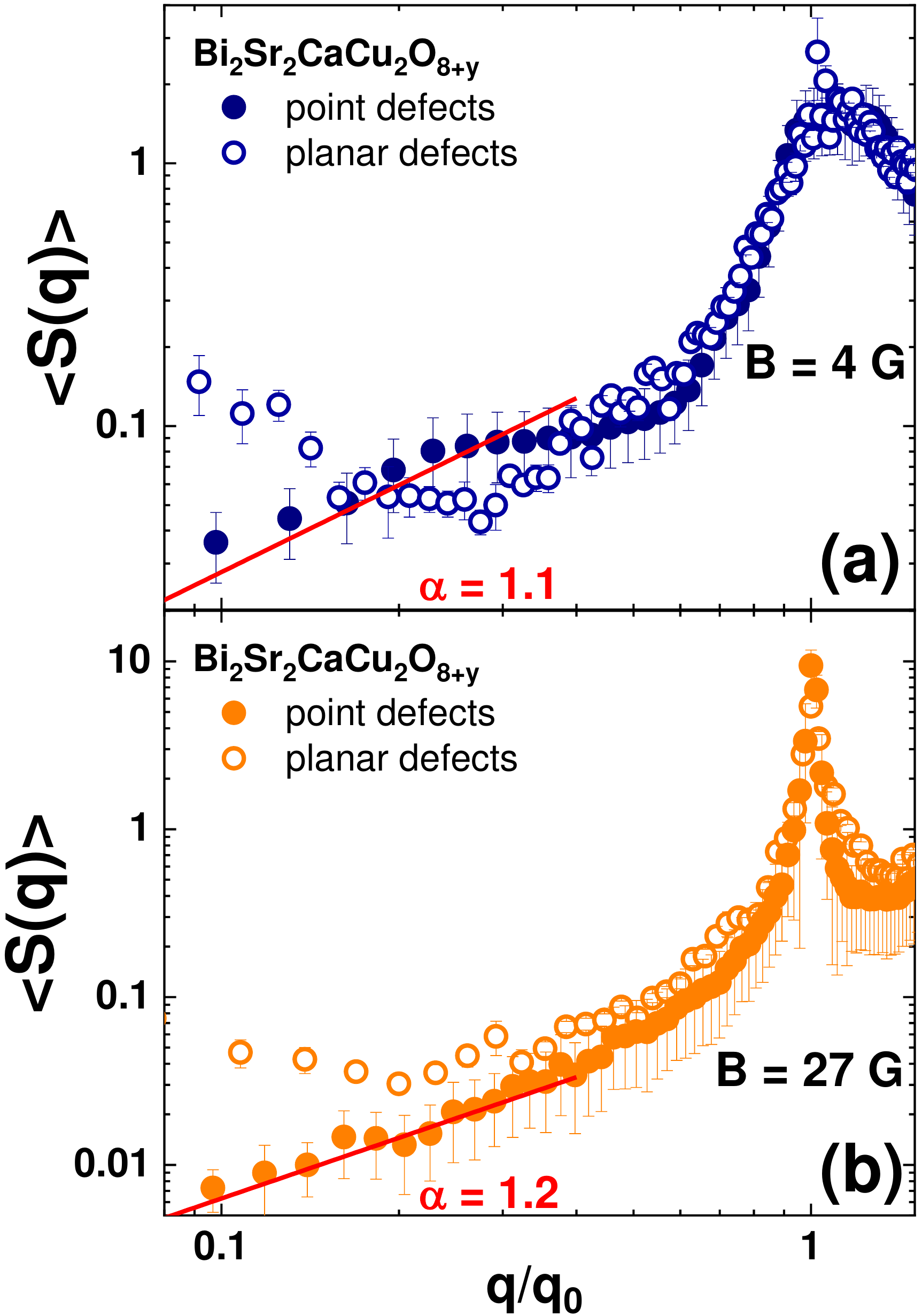}
    \caption{Angularly-averaged structure factor of the vortex structures nucleated
        in Bi$_2$Sr$_2$CaCu$_2$O$_{8+\delta}$ with point (full symbols) and planar
        (open symbols) disorder for densities of (a) 4 and (b) 27\,G. The wavenumber
        is normalized by the Bragg wavenumber $q_{0}=2\pi/(a_{0} \cos{\pi/6}$. Full red lines are fits
        to the data in samples with point disorder considering $\langle
S(q)\rangle \sim (q/q_{0})^{\alpha}$ when $q/q_{0} \rightarrow 0$
yielding the $\alpha$ exponents indicated in each case. Data  for
the 27\,G structures are obtained considering the position of 6189
(2100) vortices in a sample pristine (with planar defects), whereas
data for the 4\,G structures come from 500 (2170) vortices for a
sample pristine (with planar defects).} \label{fig:Figure2}
\end{figure}

In order to quantify how a softening of the structure affects the
density fluctuations at long wavelengths, we calculated the
structure factor and then computed its angular-average at every $q$,
$\langle S(\mathbf{q}) \rangle$. The result of this analysis is
shown in Fig.\,\ref{fig:Figure2} for both vortex densities, plotted
as a function of $q/q_{0}$ with $q_{0}=2\pi/(a_{0} \cos{\pi/6}$ the
Bragg wavevector of the hexagonal vortex structure. Data in full
symbols correspond to the case of point disorder whereas open
symbols come from structures nucleated in samples with planar
defects. In the first case, $\langle S(\mathbf{q}) \rangle$ decays
algebraically when $q \rightarrow 0$ with the exponents $\alpha=1.1$
for 4\,G and 1.2 for 27\,G. In both cases these exponents are
yielded by fits of the data in the 0.1-0.4\,$q/q_{0}$ range with an
error of 0.3 for 4\,G and 0.2 for 27\,G. Thus, as already reported
on Ref.\,\cite{Rumi2019}, density fluctuations in samples with point
disorder decay at long wavelengths in a hyperuniform fashion
irrespective of the vortex lattice softness.

In contrast, in the case of samples with planar defects the
angularly-averaged structure factor saturates and even has a
tendency to unboundedness when $q \rightarrow 0$. This indicates
that vortex density fluctuations do not tend to zero at long
wavelengths but they rather enhance with distance. Thus, in samples
with correlated disorder the vortex structure does not present the
hyperuniform hidden order and is instead anti-hyperuniform. This
phenomenology takes place even for the harder vortex structure of
27\,G with a larger vortex-vortex interaction, a case where the
relative effect of the pinning induced by planar defects is smaller.
Nevertheless, $\langle S(\mathbf{q}) \rangle$ grows faster as $q
\rightarrow 0$ when the softness of the structure increases. This
indicates that for softer vortex structures the ability of planar
defects to suppress hyperuniformity is enhanced. Whether this is
related to a different relevance of the finite-size effect that may
lead to a recovery of hyperuniformity in samples with planar defects
that are sufficiently thick as reported in Ref.\,\cite{Puig2022},
deserves further theoretical investigation.

\section{Conclusions}

Here we report that planar correlated disorder is effective to
enhance long wavelength density fluctuations as to suppress the
hyperuniform hidden order otherwise present in the vortex structure
nucleated in samples with point disorder. This effect takes place
irrespective of the magnitude of vortex-vortex interaction but it is
enhanced in the case of softer structures of interacting elastic
objects such as vortices in type-II superconductors.


\begin{thebibliography}{99}

\bibitem{Rumi2019} G. Rumi, J. Arag\'on S\'anchez, J., F. El\'{i}as,
R. Cort\'es Maldonado, J. Puig, N. R. Cejas Bolecek, G. Nieva,
M. Konczykowski, Y. Fasano, and A. B. Kolton, Phys. Rev. Res.
\textbf{1}, 033057 (2019).


\bibitem{Llorens2020b} J. B. Llorens, I. Guillam\'on, I. Garc\'{i}a-Serrano, R. C\'ordoba, J. Ses\'e, J. M. De Teresa, M. R. Ibarra, S. Vieira, M. Ortu\~{n}o, and H. Suderow, Phys. Rev. Res. \textbf{2}, 033133 (2020).


\bibitem{Man2013}  W. Man, M. Florescu,, E. P.    Williamson, Y.  He, S. R.  Hashemizad, B. Y. C.  Leung, D. R.  Liner, S. Torquato, P. M. Chaikin, and P. J. Steinhardt, Proceed. Nat. Acad. Sci. USA \textbf{110}, 15886 (2013).


\bibitem{Chen2018} D. Chen, and S. Torquato, Acta Materialia \textbf{142}, 152 (2018).

\bibitem{Torquato2018} S. Torquato, Phys. Rep.
\textbf{745},  \textbf{1} (2018).

\bibitem{Zheng2020} Y. Zheng, L. Liu, H. Nan, Z.-X. Shen, G. Zhang, D. Chen, L. He, W. Xu, M. Chen, Y. Jiao, H. Zhuang, Sci. Adv. \textbf{6}, eaba0826 (2020).

\bibitem{Florescu2009} M. Florescu, S. Torquato, and P. J. Steinhardt,  Proceed. Nat. Acad. Sci. USA \textbf{106}, 20658 (2009).


\bibitem{FroufePerez2016} L. S. Froufe-P\'{e}rez,  M. Engel, P. F. Damasceno, N. Muller, J. Haberko, S. C. Glotzer, and F. Scheffold, Phys. Rev. Lett. \textbf{117}, 053902 (2016).

\bibitem{Torquato2003} S. Torquato, and F. H.  Stillinger,  Phys. Rev. E \textbf{68}, 041113 (2003).

\bibitem{Klatt2019} M. A. Klatt, J. Lovric, D. Chen, S. C. Kapfer, F. M. Schaller, P. W. A. Sch\"{o}nh\"{o}fer, B. S. Gardiner, A.-S. Smith, G. E. Schr\"{o}der-Turk, and S. Torquato, Nature Comm. \textbf{10}, 811 (2019).

\bibitem{Aragon2022} J. Arag\'{o}n S\'{a}nchez, R. Cort\'{e}s Maldonado, L. Amig\'{o}, G. Nieva, A. B. Kolton, and Y. Fasano, submitted to
 Phys. Rev. Lett. (2022)


\bibitem{Puig2021} J. Puig, F. El\'{i}as, J. Arag\'{o}n S\'{a}nchez, R. Cort\'{e}s Maldonado, G. Rumi, G. Nieva, P. Pedrazzini, A. B. Kolton, and Y. Fasano, Commun. Mater. \textbf{3}, 32 (2022).

\bibitem{Fasano2008} Y. Fasano, and M. Menghini,  Supercond. Sci. Tech. \textbf{21},
 023001 (2008).


\bibitem{Fasano2003} Y. Fasano, M. De Seta, M. Menghini, H. Pastoriza, and F. de la Cruz,  Solid
State Comm. \textbf{128}, 51 (2003).


\bibitem{CejasBolecek2016}  N. R. Cejas Bolecek, A. B. Kolton, M. Konczykowski, H. Pastoriza, D. Dom\'{i}nguez, and Y. Fasano,  Phys. Rev. B \textbf{93}, 054505
(2016).


\bibitem{Pardo1997}  F. Pardo, A. Mackenzie, F.  de la Cruz, and J. Guimpel,
 Phys. Rev. B \textbf{55}, 14610 (1997).

 \bibitem{Fasano1999}  Y. Fasano, J. A.  Herbsommer, F. de la Cruz, F. Pardo, P. L. Gammel,  E. Bucher, and D. J. Bishop, Phys. Rev. B \textbf{60}, 15047 (1999).

\bibitem{Correa2001} V. F. Correa, E. E. Kaul, and G. Nieva, Phys. Rev. B \textbf{63}, 172505 (2001).

\bibitem{Koblischka1995}
M. R. Koblischka, \textit{ et al.}, Phys. C 249, 339 (1995).

\bibitem{Herbsommer2001}  J. A. Herbsommer, V. F. Correa, G. Nieva, H. Pastoriza, and J. Luzuriaga, Solid State
Comm. \textbf{120}, 59 (2001).




\end{thebibliography}
\end{document}